\documentclass[twocolumn]{article}

\usepackage[modulo,switch]{lineno}
\modulolinenumbers[1]

\usepackage[latin9]{inputenc}
\usepackage{amsfonts}
\usepackage{amsmath}
\usepackage{amssymb}
\usepackage{bm}
\usepackage{graphicx}
\usepackage{color}
\usepackage{cite}

\makeatletter\@ifundefined{date}{}{\date{}}
\makeatother

\evensidemargin\oddsidemargin \hoffset-22.4mm \voffset-28.4mm
\newcommand{\bs}{\boldsymbol}
\newcommand{\eff}{\mathrm{eff}}

\begin{document}

\title{On the construction of multiscale surrogates for design optimization of acoustical materials}

\author{Van Hai Trinh$^{1,2)}$, Johann Guilleminot$^{3)}$, Camille Perrot$^{1)}$\\
$^{1)}$ Université Paris-Est, Laboratoire Modélisation et Simulation Multi Echelle\\
MSME UMR 8208 CNRS, Marne-La-Vallée 77454, France. camille.perrot@u-pem.fr\\
$^{2)}$ Le Quy Don Technical University, Hanoi, Vietnam\\
$^{3)}$ Department of Civil and Environmental Engineering, Duke University\\
Durham, NC 27708, USA.}

\maketitle\thispagestyle{empty}

\begin{abstract}
This paper is concerned with the use of polynomial metamodels for the design of acoustical materials, considered as equivalent fluids. Polynomial series in microstructural parameters are considered, and allow us to approximate the multiscale solution map in some well-defined sense. The relevance of the framework is illustrated by considering the prediction of the sound absorption coefficient. In accordance with theoretical results provided elsewhere in the literature, it is shown that the surrogate model can accurately approximate the solution map at a reasonable computational cost, depending on the dimension of the input parameter space. Microstructural and process optimization by design are two envisioned applications.
\end{abstract}

\section{Introduction}\label{sec:intro}
The inverse design of materials has recently gained popularity in both academia and industry. Materials by design approaches typically require (i) the construction of a mapping between the microstructural features at some relevant scale and the properties of interest (with a desired level of accuracy), and (ii) the design of an optimization algorithm that can efficiently explore innovative solutions. In this paper, we investigate the use of a multiscale-informed polynomial surrogate to define an approximation of the macroscopic acoustical properties in terms of microstructural variables.

Let $\bs{m}$ denote the vector of microstructural parameters to be optimized, and assume that $\bs{m}$ belongs to some admissible closed set $\mathcal{S}_{\bs{m}} = \times_{i = 1}^{n} [a_i,b_i]$ in $\mathbb{R}^n$. Let $\bs{q} \in \mathcal{S}_{\bs{q}} \subseteq \mathbb{R}^d$ be some macroscopic quantity of interest. Microstructural design optimization then consists in finding, using an \textit{ad hoc} computational strategy, the optimal value $\bs{m}^{opt}$ of $\bs{m}$ (which may be non-unique) minimizing some application-dependent cost function $J$ such that $J(\bs{q}) = J(\bs{q}(\bs{m})) =: J(\bs{m})$, by an abuse of notation:
\begin{equation}\label{eq:def-optimization}
\bs{m}^{opt} = \underset{\bs{m}\,\in\,\mathcal{S}_{\bs{m}}}{\mathrm{argmin}}~ J(\bs{m})~.
\end{equation}
In practice, solving the above optimization problem (which is not convex and may exhibit many local minima) requires performing multiscale simulations a large number of times, especially for large values of $n$. A classical remedy to this computational burden relies on the construction of a surrogate mapping $\bs{\hat{q}}$ that properly approximates $\bs{q}$ (that is, the map $\bs{m} \mapsto \bs{\hat{q}}(\bs{m})$ approaches the solution map $\bs{m} \mapsto \bs{q}(\bs{m})$ in some sense) and remains much cheaper to evaluate than full-field upscaling simulations. Available techniques include the use of neural networks, response surfaces \cite{1} and reduced-order models \cite{2}. Once the approximation has been defined, the optimal solution is then defined as
\begin{equation}\label{eq:def-optimization-surr}
\bs{m}^{opt} = \underset{\bs{m}\,\in\,\mathcal{S}_{\bs{m}}}{\mathrm{argmin}}~ \hat{J}(\bs{m})~, \quad \hat{J}(\bs{m}) = J(\bs{\hat{q}}(\bs{m}))~.
\end{equation}

\section{Methodological aspects}
The definition of the surrogate model $\bs{\hat{q}}$ involves key theoretical questions (such as the characterization of convergence rates), as well as algorithmic concerns (related to the design of efficient strategies to build the metamodel, for instance). These issues have attracted much attention in various fields, especially for the computational treatment of partial differential equations, and an extensive review on this topic is beyond the scope of this letter (see e.g., \cite{2} for a survey, as well as \cite{3,4} and the references therein for convergence results). Despite this fact, the use of metamodeling remains quite unexplored in the multiscale analysis of acoustic properties. Since the reference map $\bs{m} \mapsto \bs{q}(\bs{m})$ typically introduces some smoothness due to its multiscale nature, polynomial approximation techniques are natural candidates for the construction of $\bs{\hat{q}}$ (see e.g., \cite{4}). Upon introducing the normalized vector-valued parameter $\bs{\xi}$ such that $[-1,1] \ni \xi_i = 2/(b_i-a_i)m_i + (a_i+b_i)/(a_i-b_i)$ for $1 \leqslant i \leqslant n$, the surrogate model $\bs{\hat{q}}$ is then sought for as a polynomial map in $\bs{\xi}$:
\begin{equation}\label{eq:def-q-hat}
\bs{\hat{q}}(\bs{\xi}) = \sum_{\bs{\alpha}} \bs{\hat{q}}_{\bs{\alpha}} P_{\bs{\alpha}}(\bs{\xi})~,
\end{equation}
where $\bs{\alpha}$ is a multi-index in $\mathbb{N}^n$, $P_{\bs{\alpha}}$ is the multidimensional Legendre polynomial defined as $P_{\bs{\alpha}}(\bs{\xi}) = \prod_{i = 1}^{n} P_{\alpha_i}(\xi_i)$, and $P_{\alpha_i}$ is the univariate Legendre polynomial of order $\alpha_i$ (see e.g., Chapter 8 in \cite{4p}). From the orthogonality of these polynomials, namely
\begin{equation}
\begin{aligned}
<P_{\bs{\alpha}}, P_{\bs{\beta}}> &=  \frac{1}{2^n} \int_{([-1,1])^n} P_{\bs{\alpha}}(\bs{x})\, P_{\bs{\beta}}(\bs{x})\,d\bs{x} \\
& = \prod_{i=1}^{n}\frac{\delta_{\alpha_i \beta_i}}{2\alpha_i+1}~,
\end{aligned}
\end{equation}
where $\delta$ is the Kronecker delta, it follows that
\begin{equation}\label{eq:def-coef}
\bs{\hat{q}}_{\bs{\alpha}} = \left(\prod_{i=1}^{n} (2\alpha_i+1) \right) <\bs{\hat{q}},P_{\bs{\alpha}}>~.
\end{equation}
The choice of this polynomial basis ensures that the surrogate is uniformly accurate over the parameter space, so that no bias (noise) is generated in the evaluation of the cost function. The computation of the coefficients $\bs{\hat{q}}_{\bs{\alpha}}$ necessitates the evaluation of $n$-dimensional integrals, and various techniques have been proposed in the literature to address this issue. Standard or enhanced (i.e. nested, sparse, etc.) quadrature rules can be invoked for small values of $n$, while (advanced) Monte Carlo simulation techniques can be used for much higher dimensions (see e.g., \cite{5}). Below, a Gauss-Legendre quadrature rule is used for illustration purposes.

\section{Numerical results}
\subsection{Reference solution map}
\begin{figure}[ht]
\centering \includegraphics[width=0.4\textwidth,trim={0 4cm 0 3.75cm},clip]{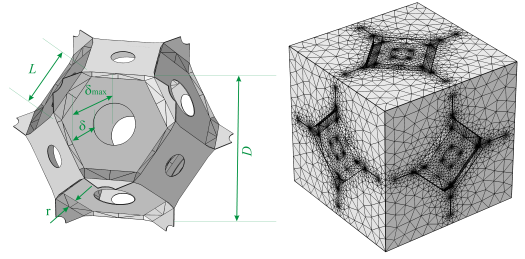}
\caption{\label{fig:FIG1}{Unit cell and FE mesh ($\phi = 0.97$, $r_c = 0.6$).}}
\end{figure}

\begin{figure}[!h]
\centering \includegraphics[width=0.4\textwidth]{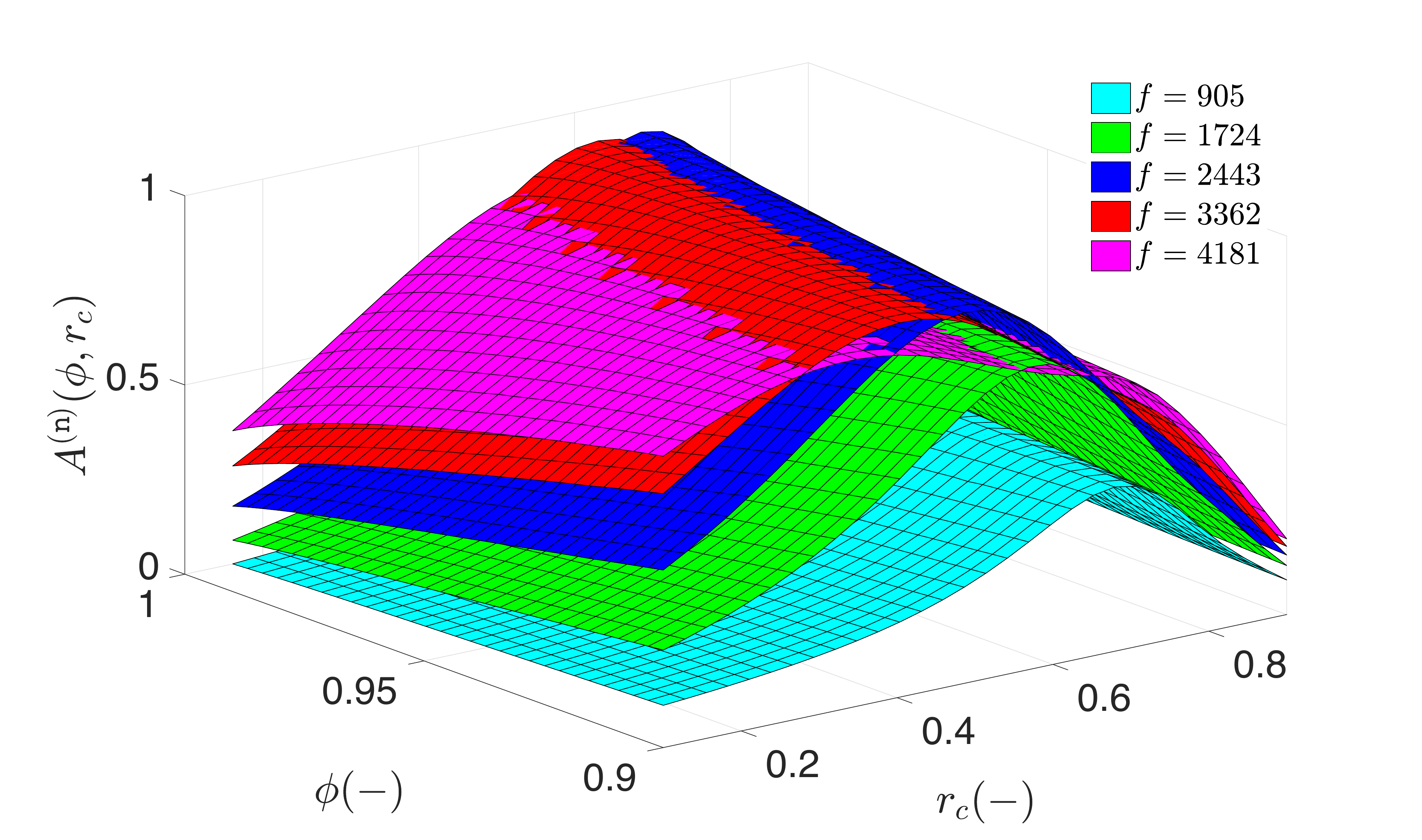}
\caption{\label{fig:FIG2}{Reference solution map $(\phi,r_c) \mapsto A^{(\mathsf{n})}(\phi,r_c)$ at different frequencies (in Hz).}}
\end{figure}

In the sequel, we consider the optimization of a tetrakaidecahedron structure (see Fig.~\ref{fig:FIG1}) for sound absorption purposes, and seek an approximation of the normal incidence sound absorption coefficient $A^{(\mathsf{n})}$ as a function of both the macroscopic porosity $\phi$ and the membrane closure rate $r_c = \delta/\delta_\mathrm{max}$. For later use, let $A^{(\mathsf{d})}$ be the sound absorption coefficient for a diffuse field excitation (see Eqs.~(7--9) in \cite{12}). Note that in a more general setting, the interpolation of intrinsic parameters, such as transport properties, is more appropriate, since they constitute primary variables enabling the prediction of e.g., frequency dependent response functions. Depending on the context, $A^{(\mathsf{n})}$ is indexed by either the frequency $f$ or the angular frequency $\omega = 2\pi f$. We then adopt the notation $A^{(\mathsf{n})}(\phi, r_c; f)$ (or $A^{(\mathsf{n})}(\phi, r_c; \omega)$), and any variable temporarily fixed may be dropped with no notational change (when $\phi$ and $r_c$ are fixed, the absorption coefficient simply reads as $A^{(\mathsf{n})}(f)$ or $A^{(\mathsf{n})}(\omega)$). While changes in the porosity $\phi$ can be imposed in various ways, we presently consider adapting the ligament thickness $r$ (as shown Fig.~\ref{fig:FIG1}) and the size $D$ of the unit cell remains constant and equal to $0.8$ mm.
Furthermore, the same closure rate is imposed on all faces of the structure, which reflects both the assumed periodicity and processing constraints. Following the notations introduced in \S~\ref{sec:intro}, $\bs{m}$ is identified with the vector $(\phi, r_c)$ and $\bs{q} = (A^{(\mathsf{n})})$; hence, $n = 2$ and $d = 1$. For a given value of the microstructural parameters, $A^{(\mathsf{n})}(\omega)$ is obtained as $A^{(\mathsf{n})}(\omega) = 1-\left|(Z_s(\omega) - Z_0)/(Z_s(\omega) + Z_0)\right|^2$, where $Z_0$ is the air impedance and $Z_s(\omega)$ is the normal incidence surface impedance of the equivalent fluid. For a layer of thickness $L_s$ ($L_s = 20$ mm below), $Z_s(\omega)$ reads as $Z_s(\omega) = - j Z_c(\omega) \cot(k_c(\omega)L_s)$, where $j$ is the imaginary unit, $Z_c(\omega)$ is the characteristic impedance and $k_c(\omega)$ denotes the wave number (with the time convention: $+ j \omega t$). These quantities can be expressed in terms of the effective density $\rho_\eff(\omega)$ and effective bulk modulus $K_\eff(\omega)$ as $Z_c(\omega) = \sqrt{\rho_\eff(\omega)K_\eff(\omega)}$ and $k_c(\omega) = \omega \sqrt{\rho_\eff(\omega)/K_\eff(\omega)}$. The effective properties can be estimated by using the semi-phenomenological JCAPL model \cite{6,7,8,9}, which involves transport properties that are obtained by solving a set of independent boundary value problems (BVPs) (Stokes, potential flow and thermal conduction equations; see e.g., Chapter 5 in \cite{10} and  Appendix B in \cite{11} for a condensed presentation of this model). In this work, these BVPs are solved by using the finite element method (at convergence, the mesh associated with the complete cell contains $214,412$ tetrahedral elements; see Fig.~\ref{fig:FIG1}) and the commercial software COMSOL Multiphysics. For a given configuration (i.e. for given values of $\phi$ and $r_c$), the averaged computation time for the multiscale simulations is about $156$ seconds. The reference solution map is shown in Fig.~\ref{fig:FIG2} for various frequencies.

\subsection{Surrogate analysis}
\begin{figure}[!h]
\centering \includegraphics[width=0.4\textwidth]{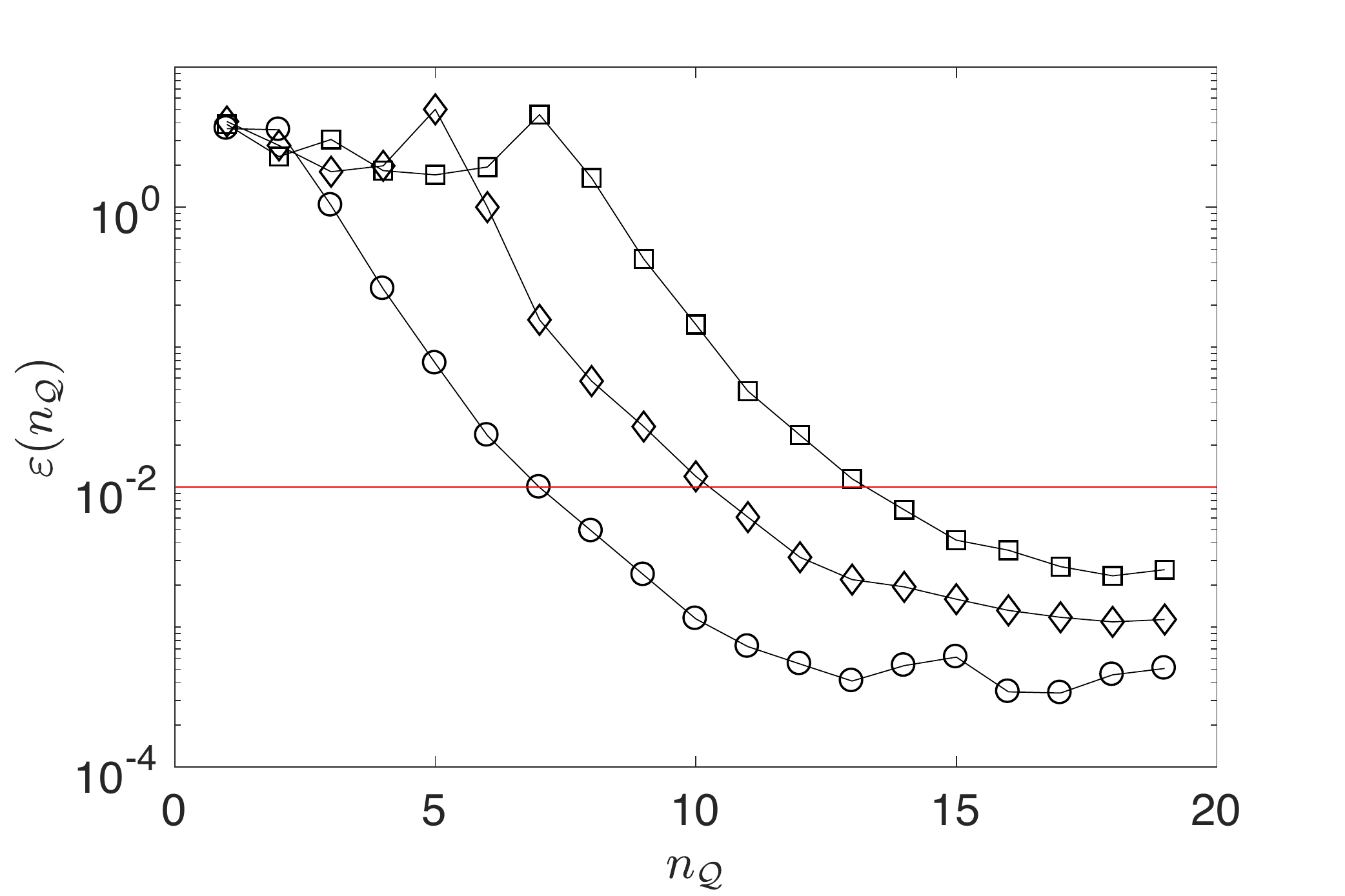}
\caption{\label{fig:FIG6}{Graph of the error function $n_{\mathcal{Q}} \mapsto \varepsilon(n_{\mathcal{Q}})$ for $p = 5$ (circles), $10$ (diamonds) and $15$ (squares).}}
\end{figure}

It follows from Eq.~(\ref{eq:def-q-hat}) that the approximant, truncated at order $p$, is given by
\begin{equation}
\bs{\hat{q}}_p(\bs{\xi}) = \sum_{\bs{\alpha}\,\in\, \mathbb{N}^2, \; |\bs{\alpha}| = 0}^{p} \bs{\hat{q}}_{\bs{\alpha}} P_{\bs{\alpha}}(\bs{\xi})~,
\end{equation}
where $\bs{\xi} = (\xi_1, \xi_2)$, $|\bs{\alpha}| = \sum_{i = 1}^{n} \alpha_i$ and $P_{\bs{\alpha}}(\bs{\xi}) = P_{\alpha_1}(\xi_1) P_{\alpha_2}(\xi_2)$. The reduced coordinate $\xi_1$ corresponds to $\phi$, and $\xi_2$ represents the closure rate $r_c$. The coefficients are then estimated (see Eq.~\ref{eq:def-coef}) as
\begin{equation}\label{eq:comput-coef}
\bs{\hat{q}}_{\bs{\alpha}} \approx \frac{(2\alpha_1+1)(2\alpha_2+1)}{4} \sum_{i = 1}^{N_{\mathcal{Q}}} w_i \, \bs{\hat{q}}(\bs{\xi}^{(i)}) P_{\bs{\alpha}}(\bs{\xi}^{(i)})~,
\end{equation}
in which $\{w_i\}_{i = 1}^{N_{\mathcal{Q}}}$ and $\{\bs{\xi}^{(i)}\}_{i = 1}^{N_{\mathcal{Q}}}$ are the weights and points of the quadrature rule. Evaluating the multiscale model at the quadrature points represents offline stage (distributed) computations in which the reduced variables are mapped back onto the physical ones (i.e. $\phi$ and $r_c$). Convergence must be characterized with respect to both $p$ (using e.g., a $L^2$ metric for increasing orders of expansion) and $n_{\mathcal{Q}} = (N_{\mathcal{Q}})^{1/2}$ (for a fixed order of expansion $p$). In practice, the value of $n_{\mathcal{Q}}$ can be determined by analysing the convergence of the function $n_{\mathcal{Q}} \mapsto \varepsilon(n_{\mathcal{Q}})= \|\bs{\hat{q}}_{\bs{\alpha}}(n_{\mathcal{Q}}) - \bs{\hat{q}}_{\bs{\alpha}}(n_{\mathcal{Q}}+1)\|^2/\|\bs{\hat{q}}_{\bs{\alpha}}(n_{\mathcal{Q}})\|^2$, where the dependence of $\bs{\hat{q}}_{\bs{\alpha}}$ on $n_{\mathcal{Q}}$ is made explicit (see Eq.~(\ref{eq:comput-coef})). In what follows, $n_{\mathcal{Q}}$ is determined such that $\varepsilon(n_{\mathcal{Q}}) \leqslant 10^{-2}$ (see Fig.~\ref{fig:FIG6}).

Let $\mathcal{D}_p$ be the relative error measure defined as $\mathcal{D}_p(\phi,r_c)= |A^{(\mathsf{n})}(\phi,r_c) - \hat{A}_p^{(\mathsf{n})}(\phi,r_c)|/A^{(\mathsf{n})}(\phi,r_c)$, where $\hat{A}_p^{(\mathsf{n})}$ is the estimate of the sound absorption coefficient (normal incidence) obtained with the surrogate model at order $p$. The probability density function of $\mathcal{D}_p$ obtained for $\phi \in [0.9,0.99]$ and $r_c \in [0.1, 0.9]$ (with a total of $900$ combinations evaluated) is shown in Fig.~\ref{fig:FIG3}, for $p = 15$ (with $n_\mathcal{Q} = 14$, implying that $196$ computations are necessary to calibrate the surrogate model).
\begin{figure}[!h]
\centering \includegraphics[width=0.4\textwidth]{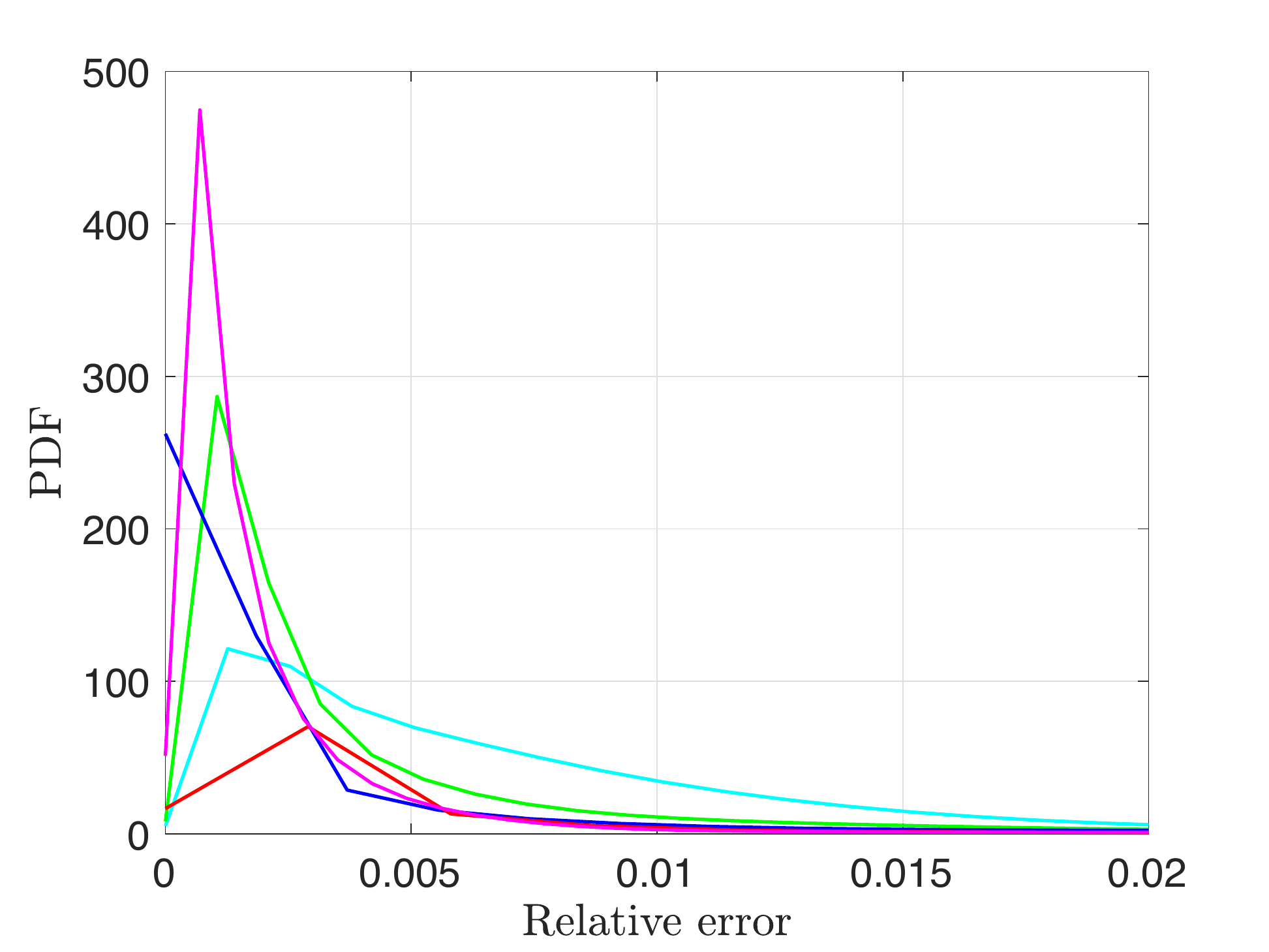}
\caption{\label{fig:FIG3}{PDF of the relative error for $p = 15$ and for the set of frequencies shown in Fig.~\ref{fig:FIG2}.}}
\end{figure}
As expected, uniform convergence over the parameter space is observed, with a relative error that is typically less than $2\%$, regardless of the frequency under consideration. It should be noticed that the apparent ordering in mean and variance, which both decrease when the frequency increases, is due to the frequency dependency of the normalizing absorption coefficient (see Fig.~\ref{fig:FIG2}). The accuracy of the approximation can also be assessed over a wide range of frequencies, as shown in Fig.~\ref{fig:FIG3} for $p = 10$ ($n_\mathcal{Q} = 11$) and $p = 15$ ($n_\mathcal{Q} = 14$).
\begin{figure}[!h]
\includegraphics[width=0.5\textwidth]{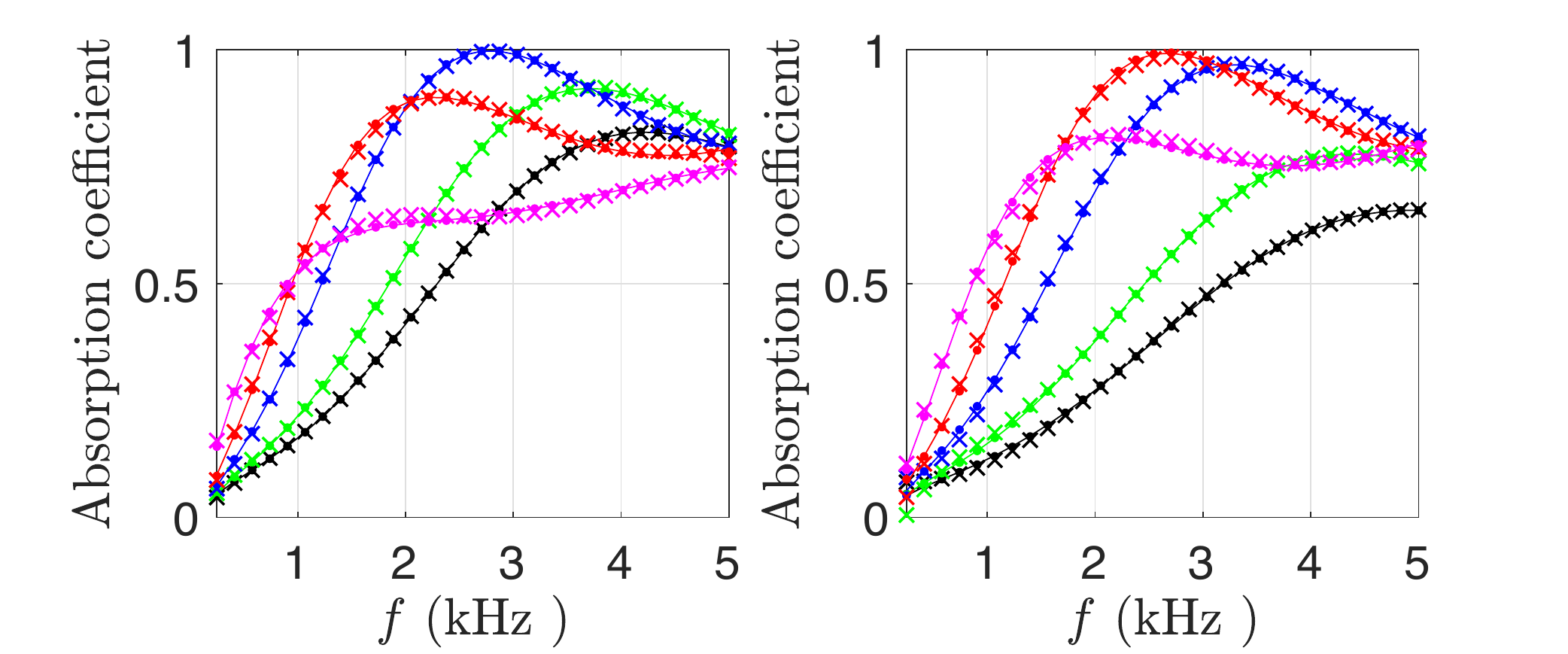}
\caption{\label{fig:FIG3}{Solution map for the normal incidence sound absorption coefficient. Solid line: reference; cross markers: surrogate with $p = 10$; point markers: surrogate with $p = 15$. The results are shown for $r_c = 0.1$ (black), $0.2931$ (green), $0.4862$ (blue), $0.5966$ (red), $0.7069$ (magenta), and $\phi=0.9124$ (left panel) and $\phi=0.9745$ (right panel).}}
\end{figure}

\begin{figure}[!h]
\centering \includegraphics[width=0.45\textwidth, trim={1cm 0cm 1cm 0cm},clip]{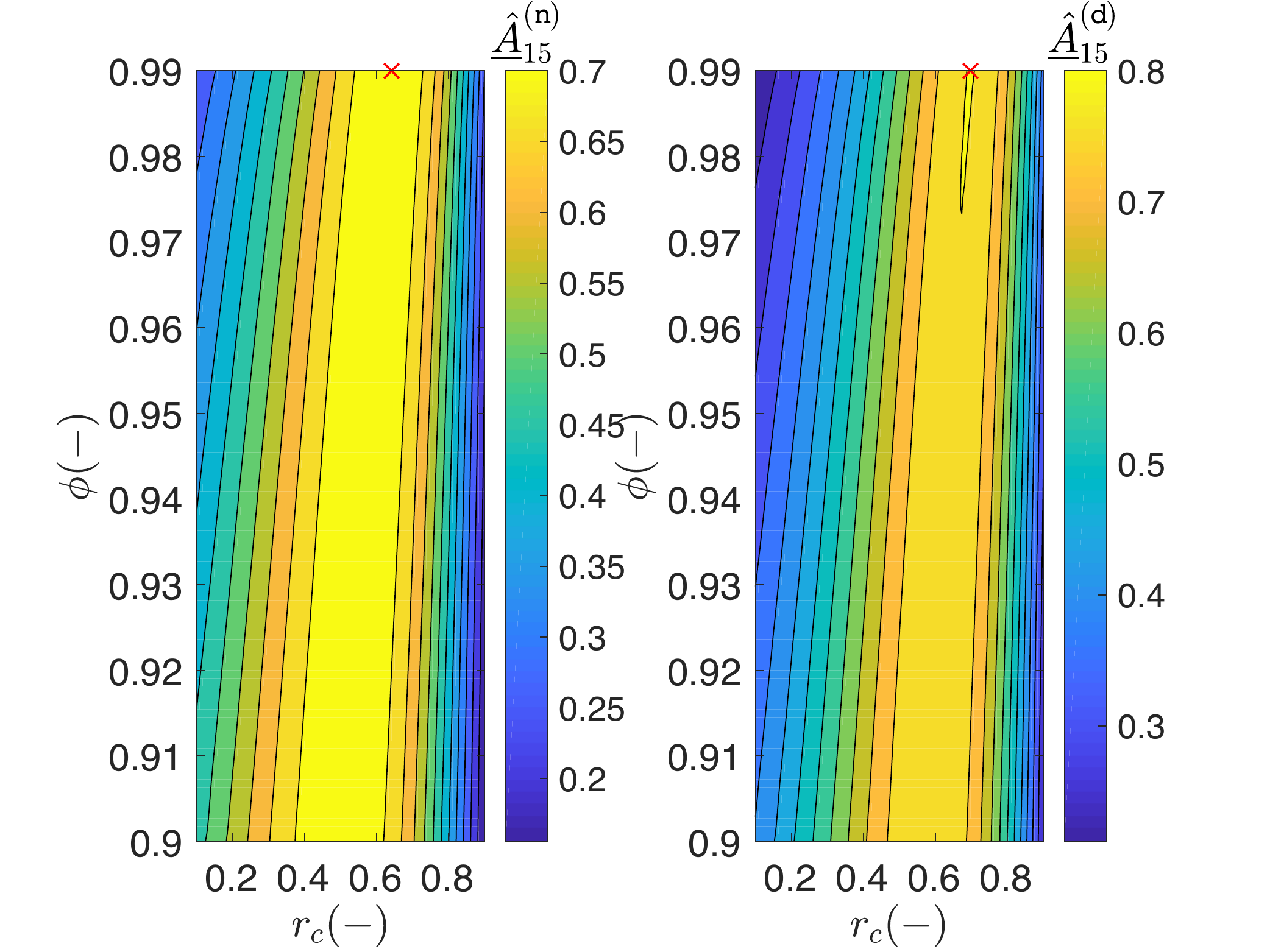}
\caption{\label{fig:FIG4}{Plots of the averaged absorption coefficients, with $f_0 = 250$ and $f_1 = 5,000$ Hz. The maximum value in each chart is identified with a red cross.}}
\end{figure}

Let us now consider the optimization problem given by Eq.~(\ref{eq:def-optimization-surr}), and consider, for $\bs{m} = (\phi,r_c)$, the cost function $\hat{J}(\bs{m}) = - \bs{\hat{q}}_\beta(\bs{m})$, with $\beta \in [0,1]$ and
\begin{equation}\label{eq:def-hat-q-example}
\bs{\hat{q}}_\beta(\bs{m}) = \beta \underline{\hat{A}}_p^{(\mathsf{n})}(\bs{m}) + (1-\beta) \underline{\hat{A}}_p^{(\mathsf{d})}(\bs{m})
\end{equation}
where $\underline{\hat{A}}_p^{(\mathsf{n})}(\bs{m})$ and $\underline{\hat{A}}_p^{(\mathsf{d})}(\bs{m})$ are the averages of the sound absorption coefficients, approximated with the surrogate, over the frequency interval $[f_0, f_1]$:
\begin{equation}
\underline{\hat{A}}_p^{(\mathsf{k})}(\bs{m})=\frac{1}{f_{1} - f_{0}} \int_{f_{0}}^{f_{1}} \hat{A}_p^{\mathsf{(k)}}(\bs{m};f)\,df~,
\end{equation}
where $\mathsf{k}$ stands either for $\mathsf{n}$ or $\mathsf{d}$. Note that the dependence of $\bs{\hat{q}}$ on $p$ is not reported to simplify notation. The charts showing the approximated sound absorption coefficients are reported in Fig.~\ref{fig:FIG4}, and can be used to evaluate the performance of the material over ranges of values induced by process variability.
Once calibrated, the surrogate model allows the cost function to be evaluated at a negligible computational expense, which opens up many possibilities to design optimal microstructures (pore size, membrane content) under contraints related to different acoustical parameters. Whereas the proposed application was concerned with transport and sound absorbing only, it should finally be noticed that the approach can readily accommodate other constraints related to mechanical and sound insulation properties in a multi-objective formulation.

\section{Conclusion}
In this work, we have investigated the potential of polynomial metamodels to accurately approximate mappings between key microstructural features and homogenized acoustical properties. The approach relies on orthogonal polynomials and enables appropriate convergence over the parameter space to be ensured. It is shown that the framework allows the sound absorption coefficient to be predicted over an appropriate range of frequencies, so that the optimization of microstructures under various types of constraints can be envisioned at a reasonable computational cost  to support the design for noise reducing materials and structures (COST Action CA15125).

\section*{Acknowledgments}
The work of V. H. Trinh was supported by a fellowship awarded by the Government of Vietnam (Project 911). The work from C. Perrot was supported in part by the French National Research Agency under grant ANR-13-RMNP-0003-01.

\small
\bibliographystyle{plain}

\end{document}